\documentclass{iau} 
\usepackage{graphicx}
\usepackage{epstopdf}

\title[Reconnection Acceleration in BHs and jets] %% give here short title %%
{Particle acceleration and the origin of the very high energy emission around \\ black holes and relativistic jets}

\author[Elisabete de Gouveia Dal Pino et al.]   %% give here short author list %%
{Elisabete  de Gouveia Dal Pino$^1$,
 Grzegorz Kowal$^2$,
 Luis Kadowaki$^1$,
 Tania E. Medina-Torrej{\'o}n$^1$,
 Yosuke Mizuno$^3$,
\and  Chandra Singh$^4$
}

\affiliation{$^1$Instituto de Astronomia, Geof{\'i}sica e Ci{\^e}ncias Atmosf{\'e}ricas (IAG-USP),\\ Universidade de S{\~a}o Paulo, R. do Mat{\~a}o, 1226 05508-090 S{\~a}o Paulo, SP Brasil
\\ email: {\tt dalpino@iag.usp.br} \\[\affilskip]
$^2$EACH, Universidade de S{\~a}o Paulo \\
$^4$Physics Institute, University of Frankfurt,
$^3$Physics Institute, University of Tel Aviv
}

\pubyear{2018}
\volume{342}  %% insert here IAU Symposium No.
\jname{Perseus in Sicily: from black hole to cluster outskirts}
\editors{Keiichi Asada, Elisabete de Gouveia Dal Pino, \\
Hiroshi Nagai, Rodrigo Nemmen, \& Marcello Giroletti, eds.}
\begin{document}

\maketitle

\begin{abstract}
Particle acceleration induced by fast magnetic reconnection 
%in the surrounds of black holes and their relativistic jets 
may help to solve current puzzles 
%specially 
related to the interpretation of the very high energy (VHE) and neutrino  emissions 
%produced in 
from AGNs and compact sources in general. 
%In this talk, we discuss this process in the context of these sources.
%, showing our recent results based on analytical and three-dimensional numerical MHD modeling  including test particles. 
Our 
%MHD and  
general relativistic-MHD  
%numerical 
simulations of accretion disk-corona systems  reveal the growth of turbulence driven by MHD instabilities that lead to the development of fast  magnetic reconnection in the corona. 
%with fast reconnection rates around $0.1 V_A$ (where $V_A$ is the local Alfv{\'e}n speed), 
%resembling  the solar corona. 
%This is  crucial for assessing  recent theories of particle acceleration and VHE emission driven by turbulence-induced fast reconnection around compact objects. 
In addition, our 
%numerical 
simulations of relativistic MHD jets 
%subject to current-driven-kink instability 
reveal the formation of several 
%localized 
sites of fast reconnection induced by current-driven kink turbulence. The injection of thousands of test particles in these 
%reconnection 
regions 
%of the jet 
%result in their 
cause acceleration up to energies of several PeVs, thus demonstrating the ability of this process to accelerate particles and produce 
%the associated 
VHE and neutrino emission, specially in blazars.  Finally, 
%in view of the results above, 
we discuss how reconnection can also explain the observed VHE luminosity-black hole mass correlation, %spanning 10 orders of magnitude, 
involving hundreds of  non-blazar sources like Perseus A,   and black hole binaries. 
\end{abstract}

\firstsection % if your document starts with a section,
\section{Introduction}
Since de Gouveia Dal Pino \& Lazarian (2005)  proposed for the first time that particles could be accelerated in a first-order  Fermi process within fast reconnection sites (current sheets), considerable progress has been attained in this field mainly boosted by numerical modeling employing both PIC (e.g. Drake et al. 2006, Zenitani \& Hoshino 2008, Lyubarsky et al. 2008, Clausen-Brown et al. 2012, Cerutti et al. 2014, Sironi et al. 2014, Guo et al. 2016) 
%\cite[]{drake06},
%\cite[]{zenitani08},
%\cite[]{lyubarsky08},
%\cite[]{clausen-brown2012}, 
%\cite[]{cerutti14}, 
%\cite[]{sironi14}, 
%\cite[]{guo16}) 
and test particle-plus-MHD techniques (e.g., Kowal, de Gouveia Dal Pino \& Lazarian 2011, 2012, del Valle, de Gouveia Dal Pino \& Kowal 2016).  
These studies  have successfully tested the theoretical predictions identifying an exponential growth of  the particle’s energy with time while trapped within current sheets (e.g., Kowal et al. 2011, 2012), and a  power-law  energy spectrum for the accelerated particles with spectral indices $\sim -1$ to $-2$ (e.g. Kowal et al. 2012, del Valle et al. 2016), and a power-law dependence of the acceleration rate with the particle energy (with power-law indices $\sim 0.3-0.6$ for a non-relativistic background (del Valle et al. 2016). These findings are general and in principle applicable to any system, specially in turbulent magnetically dominated regions.\footnote{The presence of turbulence makes magnetic reconnection naturally very fast, with reconnection velocities being a substantial fraction of the local Alfven  speed (Lazarian \& Vishniac 1999, Kowal et al. 2009).} 
This  mechanism is now regarded as important also beyond the solar system (where it has been extensively investigated). In particular, it may explain current puzzles related to very high energy phenomena observed in compact sources like AGNs, black hole X-ray binary systems, and GRBs.  In this talk, we discuss this process in the framework of these sources,  i.e., in the surrounds of black holes (BHs) and relativistic jets (see also de Gouveia Dal Pino et al. 2015, 2016 for  recent reviews).

\section{Magnetic Reconnection Acceleration in Relativistic Jets}
AGNs with highly beamed jets pointing to the line of sight, namely blazars, are the most common sources of  $\gamma-$rays. This emission is generally attributed to  diffusive shock particle acceleration along the jet  which is strongly Doppler boosted producing apparently very high ﬂuxes. However, in some cases (e.g., PKS2155-304 source; Aharonian et al. 2007), observed very high variability, of the order of minutes in the TeV range, implies  extremely compact and fast acceleration/emission regions ($< R_S /c$) with  Lorentz factors much larger than the typical bulk values ($\Gamma \sim 1-10$) expected for such sources,  in order to avoid electron-positron pair creation. So far, the only model able to explain this high variability at TeV emission seems to be magnetic reconnection involving misaligned current sheets inside the jet (e.g. Giannios et al. 2009; see also Kushwaha et al. 2017). Fast reconnection acceleration has been also invoked to explain the transition from magnetically to kinetically dominated flow and the prompt gamma-ray emission in GRBs (Zhang \& Yan 2011). 

To probe this process, we have recently performed  3D relativistic MHD simulations of rotating Poyinting flux dominated tower jets with initial helical fields (Singh, Mizuno \& de Gouveia Dal Pino 2016). Considering models with a ratio between the magnetic and the rest mass energy of the flow $\sigma  =1$,  and different density ratios between the jet and the environment, we induced precession perturbations that quickly developed current-driven kink (CDK) modes. Figure 1 (left)  shows an example of a jet with  density larger than that of the environment. The results evidence the propagation of a helical kinked structure along the jet that causes substantial dissipation of magnetic into kinetic energy. We also identify regions of maximum current density (labeled with black points) that trace filamentary current-sheets, where turbulent fast magnetic reconnection driven by the CDK  instability takes place with rates $\sim 0.05V_A$ (Singh et al. 2016; see also Kadowaki et al., in prep.). 
The right diagram of Figure 1 depicts the kinetic energy evolution of a distribution of 10,000 particles that were accelerated $in$ $situ$ within the reconnection regions of the relativistic jet depicted in the left panel. Starting with energies $ sim 1$ MeV, the particles undergo an exponential acceleration once they are trapped in the reconnection sheets (as in Kowal et al. 2102, del Valle et al. 2017), until a saturation level, when then the accelerated particles’ Larmor radius become larger than the acceleration regions. The initial background magnetic field in this simulation had a maximum value $B=0.13$ G at the jet axis, but we have also considered simulations with values up to 100 times larger. In this case, when  the particles enter the exponential regime, they  are accelerated up to energies   $\sim 10^{19}$ eV at sub-pc distances within an AGN jet (Medina-Torrejon, de Gouveia Dal Pino, Kowal, Mizuno, Singh, Kadowaki, in prep.). The implications of these results are very important for relativistic jets in general, as they indicate that these accelerated protons can produce  $\gamma-$rays in very compact regions and  also neutrinos, as recently detected  in the blazar TXS 0506+056 (IceCube collaboration et al. 2018).

\begin{figure}[b]
% \vspace*{-2.0 cm}
\begin{center}
 \includegraphics[width=5.5in]{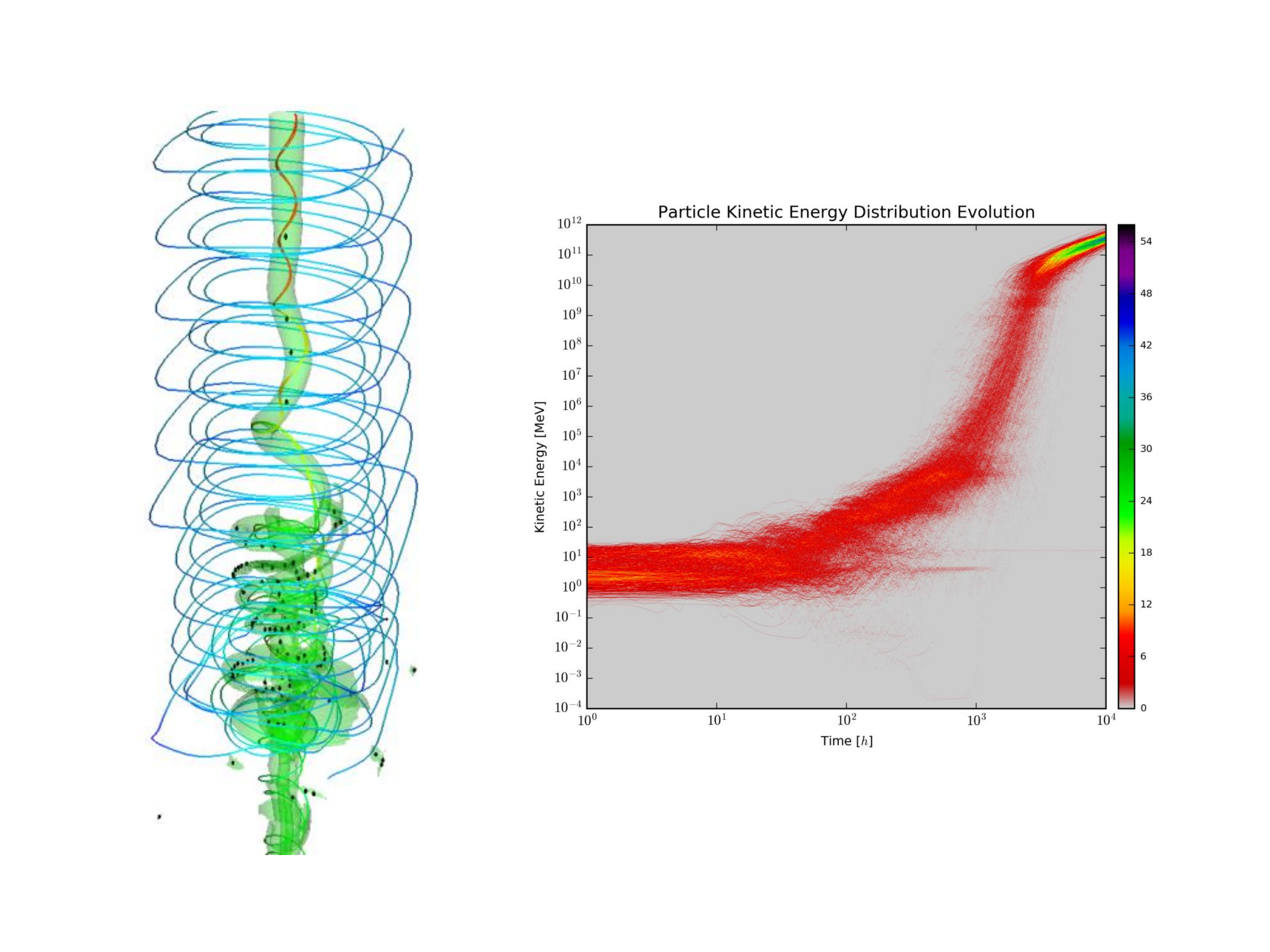} 
% \vspace*{-1.0 cm}
 \caption{Current-driven kink instability and fast magnetic reconnection in a relativistic MHD jet. The left  diagram depicts 3D density isosurfaces (green) with superposed magnetic field lines (black) (extracted from Singh et al. 2016). The black points identify the sites of fast  reconnection which were detected employing the technique described in Kadowaki, de Gouveia Dal Pino \& Stone (2018; see also Kadowaki et al., in prep.). Right  diagram shows the time evolution of the kinetic energy of 10,000 test particles (protons) that were injected with an initial Maxwelian velocity distribution in the relativistic jet of the left diagram (with initial maximum background magnetic field scaled to $B= 0.13 G$). Once particles are trapped within reconnection regions they are exponentially accelerated up to energies $\sim 10^{17}$ eV in this model. Simulations performed with magnetic field one-hundred  times larger, allow for particle acceleration up to $10^{19}$ eV  (Medina-Torrejon, de Gouveia Dal Pino, Kowal, Mizuno, Singh, Kadowaki, in prep.).}
   \label{fig1}
\end{center}
\end{figure}

\begin{figure}[b]
% \vspace*{-2.0 cm}
\begin{center}
 \includegraphics[width=5.0in]{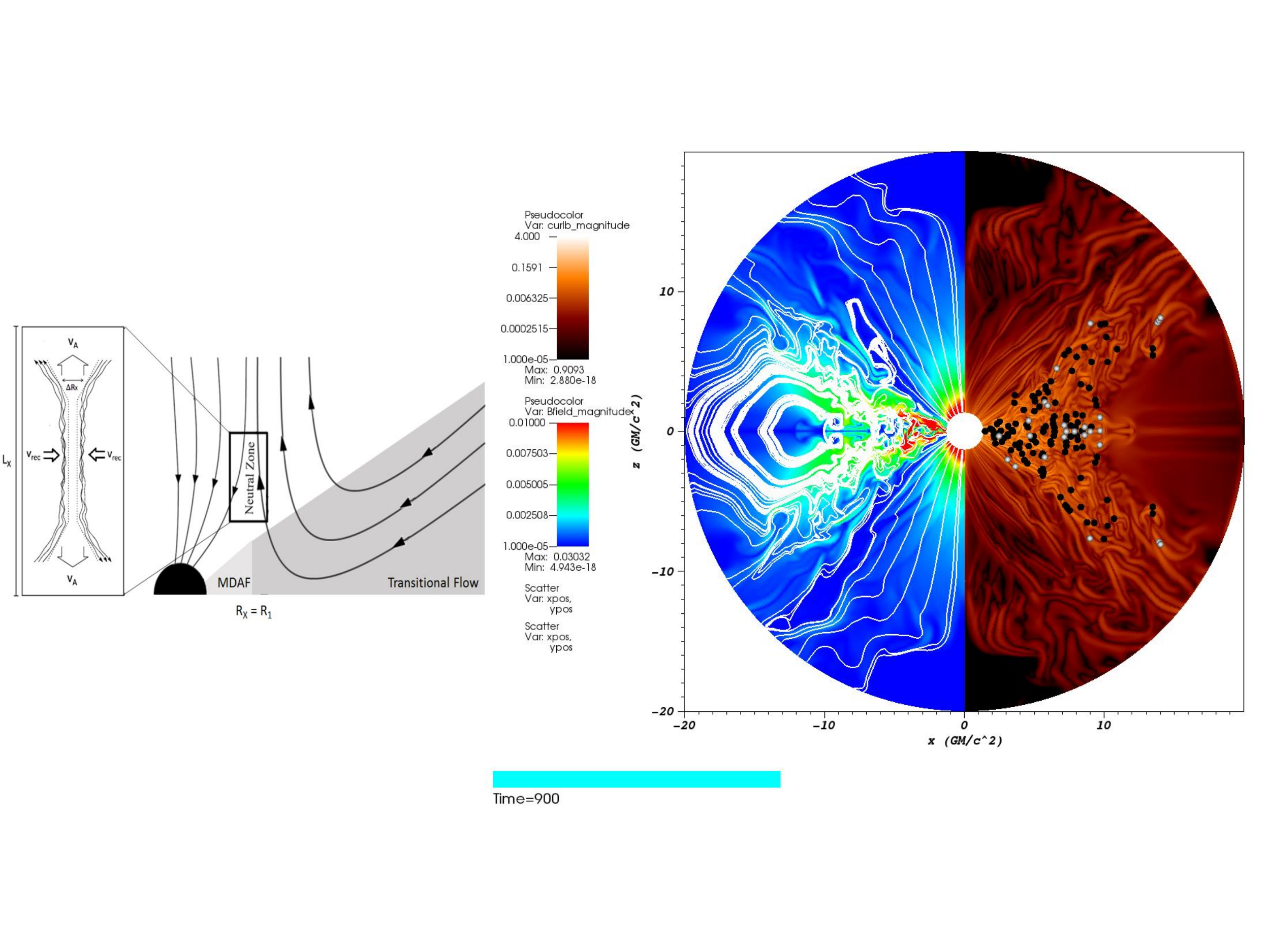} 
% \vspace*{-1.0 cm}
\caption{Left: Schematic representation of magnetic reconnection between the lines of the BH magnetosphere and those emerging from the accretion disk into the corona. Reconnection is  fast due to the presence of  turbulence in the reconnection zone (as indicated in the detail) and allows for extraction of large amounts of reconnection power. Particle acceleration may also naturally occur in the magnetic reconnection zone by  Fermi process (adapted from Singh et al. 2015). This representation is suitable for a magnetized ADAF type accretion disk,  but we obtain similar results when considering a geometrically thin accretion disk (de Gouveia Dal Pino \& Lazarian 2005, Kadowaki et al. 2015). Right: 2D general relativistic MHD (GR-MHD) simulation of a torus accretion flow around a BH evidencing the development of turbulent reconnection induced by  the MRI instability in the magnetic field and current-density distributions. The black and white dots in the current-density plot identify the sites of fast reconnection using the technique described in Kadowaki, de Gouveia Dal Pino \& Stone (2018).}
%  \label{fig2}
\end{center}
\end{figure}

\section{Magnetic reconnection acceleration around black holes} 
Recently, a few non-blazar sources which belong to the branch of low luminosity AGNs (or simply LLAGNs), have been also detected at TeV energies by ground based gamma-ray observatories (e.g.,  the radio galaxies Per A, M87,  Cen A, and IC 310; Sol et al. 2013). The angular resolution and sensitivity of these detectors are still so poor that it is hard to establish if it comes from the jet or from the core. These very high energy (VHE) detections were surprising because, besides being highly underluminous, the viewing angle of the jets of these sources is of several degrees, therefore allowing for only moderate Doppler boosting. These characteristics make it hard to explain the VHE of these sources adopting the same standard scenario as in blazars. Furthermore, observations of short time scale variability in the $\gamma-$ray emission of these sources  indicate that it is produced in a very compact region that could be, perhaps, the core. These findings led to the search for alternative particle acceleration scenarios involving the production of the VHE  in the core region of these sources (e.g. de Gouveia Dal Pino et al. 2010; 2016 and references therein). 

Another  striking evidence was found by Kadowaki, de Gouveia Dal Pino \& Singh (2015) who plotted in the same diagram the $\gamma-$ray luminosity versus the BH mass for  more than 230 sources including  non-blazar LLAGNs, black hole X-ray binaries (BHBs), blazars and GRBs, spanning over $\sim 10$ orders of magnitude in mass and power.  This diagram reveals two distinct branches, or correlation trends. One that is followed by blazars and GRBs, and another one by LLAGNs and BHBs (see Kadowaki et al. 2015, Singh, de Gouveia Dal Pino \& Kadowaki 2015). The association between blazars and GRBs is already expected as their non-thermal radiation is commonly attributed to Doppler boosted  emission by accelerated particles along the relativistic jets (as discussed in Section 2). The lack of correlation  of these sources with non-blazars and BHBs in this diagram suggests that another location and/or population of relativistic particles may be producing the $\gamma-$ray emission.   
de Gouveia Dal Pino \& Lazarian (2005)  proposed an emission model based on particle acceleration by magnetic reconnection in the core region of BH sources. This model was revisited more recently (Kadowaki et al. 2015; Singh et al. 2015) and a sketch of it is depicted in Figure 2 (left).  There is direct evidence of this process  from GR-MHD simulations of accretion disks around BHs, as indicated by  Figure 2 (right).
Fast magnetic reconnection between the lines of the BH magnetosphere and those arising from the accretion flow (with rates around 0.1 $V_A$; see Kadowaki, de Gouveia Dal Pino \& Stone in prep.) allows for the release of magnetic energy that heats and accelerates the plasma. The  magnetic  reconnection power  calculated analytically from this core model matches quite well with the observed $\gamma-$ray luminosity  of the non-blazar LLAGNs and BHBs plotted in the Luminosity-BH mass diagram described above (see Kadowaki et al. 2015; Singh et al. 2015)! This suggests that magnetic reconnection in the core region of these sources can explain the observed $\gamma-$ray emission.  Detailed calculations of the observed spectral energy distribution (SED) of a number of  BHBs (Khiali, de Gouveia Dal Pino, \& del Valle 2015) and LLAGNs, like Per A  (Ramirez-Rodriguez et al. these Procs.) using this model, have demonstrated that TeV emission in these sources (due to proton-proton and proton-photon interactions) can be effectively reproduced by accelerated cosmic rays by reconnection, and the emission is $not$ entirely absorbed by electron-positron pairs, as suspected before (Ramirez-Rodriguez, de Gouveia Dal Pino \& Alves-Batista, in prep.).  Besides, the same process leads to neutrino emission too that can account for  the observed extragalactic diffuse emission by the IceCube (see Khiali \& de Gouveia Dal Pino 2016).

\section{Summary and Conclusions}
In this talk, we have stressed the importance of turbulent magnetic reconnection and particle acceleration by reconnection  in the surrounds of BHs and relativistic jets based on arguments sustained by theory and numerical simulations. Our main conclusions can be summarized as follows:
%\item[Magnetic reconnection is important in accretion and jet systems for dissipation of magnetic energy, conversion into  kinetic energy, and particle acceleration.]	

\begin{itemize}
\item Magnetic reconnection is important in accretion and jet systems for dissipation of magnetic energy, conversion into  kinetic energy, and particle acceleration.

\item In magnetized plasmas particles can be accelerated by fast magnetic reconnection (driven by turbulence) in a Fermi process  and develop a power-law spectrum  $N(E) \sim  E^{-2,-1}$.
	  
\item Magnetic reconnection acceleration can be very efficient  in blazar jets  and produce cosmic rays with energies up to $10^{17,19}$ eV.   These could be responsible for  $\gamma-$ray flares  and neutrino emission in the magnetically dominated regions of these jets. This  could be also the dominating mechanism responsible for the recent   observed $\gamma-$ray and neutrino emission in  the TXS 0506+056 blazar and deserves further testing. 
 
\item The magnetic reconnection power can also explain the $\gamma-$ray  emission from  BHBs  and non-blazar LLAGNs  as coming from  the $core$ of these sources. This  power matches well with the observed correlation of $\gamma-$ray luminosity versus BH mass for these sources that spans 10 orders of magnitude. Accelerated CRs by magnetic reconnection  in the surrounds of these sources can produce TeV $\gamma-$rays  that are not entirely absorbed by pair production, as well as  neutrinos (see also Ramirez-Rodriguez et al. these Procs.).

\end{itemize}
 
\section{Acknowledgments}
We acknowledge support from the Brazilian agencies FAPESP (2013/10559-5 grant) and CNPq (306598/2009-4 grant). The simulations presented in this lecture have made use of the computing facilities of the GAPAE group (IAG-USP) and the Laboratory of Astroinformatics IAG/USP, NAT/Unicsul (FAPESP grant 2009/54006-4).

%\cite[Anders \& Zinner (1993)]{AndersZinner93} and 
%\cite[Ott (1993)]{Ott93}.

%\begin{discussion}

%\discuss{Massey}{I�m wondering if you have considered the expected intrinsic dispersion in absolute
%magnitude of WRs --� if you consider the (large) mass range that becomes an
%early WN or late WC according to the evolutionary models, wouldn�t you expect a large
%dispersion in M$_v$?}

%\end{discussion}

\end{document}